\documentclass[12pt,english,aps,superscriptaddress,preprintnumbers]{revtex4-1}
\usepackage[T1]{fontenc}
\usepackage[latin9]{inputenc}
\usepackage{float}
\usepackage{amsmath}
\usepackage{amssymb}
\usepackage{graphicx}
\usepackage{esint}

\makeatletter

\usepackage{mathrsfs}

\@ifundefined{textcolor}{}{%
 \definecolor{BLACK}{gray}{0}
 \definecolor{WHITE}{gray}{1}
 \definecolor{RED}{rgb}{1,0,0}
 \definecolor{GREEN}{rgb}{0,1,0}
 \definecolor{BLUE}{rgb}{0,0,1}
 \definecolor{CYAN}{cmyk}{1,0,0,0}
 \definecolor{MAGENTA}{cmyk}{0,1,0,0}
 \definecolor{YELLOW}{cmyk}{0,0,1,0}
}

\usepackage{babel}

\makeatother

\usepackage{babel}
\begin{document}
\title{Polarization of an electron scattered by static potentials}
\author{Hao-Hao Peng}
\affiliation{Shanghai United Imaging Healthcare Advanced Technology Research Institute
Co., Ltd., Shanghai 201822, China}
\author{Ren-Hong Fang}
\email{fangrh@sdu.edu.cn}

\affiliation{College of Intelligent Systems Science and Engineering, Hubei Minzu
University, Enshi, Hubei 445000, China}
\begin{abstract}
We study the polarization of an electron scattered by different static
potentials. The initial state of the electron is chosen as a wavepacket
to construct the definite orbital angular momentum, and the final
polarization of the electron, scattered by different static potentials
such as vector, pseudovector, scalar and pseudoscalar potentials,
is calculated. Numerical results show that, the sign of the polarization
of the electron scattered by the vector potential is opposite to the
other three cases, and the magnitude order of the polarization value
is consistent with recent experimental result in the collision parameter
range $0<b<2\,\mathrm{fm}$.
\end{abstract}
\maketitle

\section{Introduction}

It has long been known that particles with nonzero spin quantum number
can polarise in a rotating system \citep{Barnett:1915,Einstein:1915}.
It can be qualitatively explained by spin-orbit coupling through which
the orbital angular momentum (OAM) of the macroscopic rotating system
transfers into the spin angular momentum of microscopic particles
\citep{Liang:2004ph}. In high energy peripheral heavy-ion collisions
at STAR in BNL and at LHC in CERN, an extremely hot and dense matter
(named quark gluon plasma (QGP)) with huge OAM is created \citep{Liang:2004ph,Betz:2007kg,Becattini:2007sr,Wang:2017jpl},
hence a huge vorticity configuration is formed in this matter \citep{Becattini:2013fla,Pang:2016igs,Jiang:2016woz,Deng:2016gyh,Li:2017slc}.
It is expected that the final hadrons with nonzero spin quantum number
will polarise globally along the direction of the initial OAM at the
freeze-out stage of QGP \citep{Liang:2004ph,Gao:2007bc,Liang:2004xn,Abelev:2008ag}.
After more than ten years' effort, the global polarization of $\Lambda$
hyperon had been firstly observed at STAR \citep{Abelev:2007zk,STAR:2017ckg}.
The collisional energy dependence of the global polarization of $\Lambda$
hyperon can be well recovered by the numerical simulations of the
relativistic hydrodynamics model and the multi-phase transport model
\citep{Xie:2016fjj,Li:2017slc}. The spin alignment of vector mesons
in heavy-ion collisions are also measured these years \citep{Abelev:2008ag,Zhou:2017nwi}.

There are many methods to theoretically study the polarization of
particles in a fluid with relativistic vorticity. Based on the assumption
of local equilibrium of spin, the authors in \citep{Becattini:2013fla,Fang:2016vpj,Yang:2017sdk}
derived the relation between the 4-dimensional spin vector (Pauli-Lubanski
pseudovector) and vorticity in relativistic case through Wigner function
approach. Recently the 4-dimensional spin vector induced by vorticity
is also obtained from the method of spin density matrix but without
the factor of Fermi-Dirac thermal distribution \citep{Becattini:2019ntv}.
The first theoretical calculation of the polarization based on microscopic
scattering by a static potential is performed in \citep{Liang:2004ph},
in which the concept of global polarization of particles in high energy
heavy-ion collisions is firstly proposed. Based on \citep{Liang:2004ph},
the theoretical calculation of quark polarization through two-body
scattering \citep{Gao:2007bc} is performed, and the transport property
of polarization in a fluid is also discussed in \citep{Huang:2011ru}.
The statistical model and quark coalescense model of hydron polarization
are carried out in recent studies \citep{Liang:2004xn,Becattini:2016gvu,Yang:2017sdk,Sheng:2019kmk,Sheng:2020ghv}.
Recently one of us and his collaborators put forward a systematic
formulism to calculate the particle polarization in QGP from the spin-orbit
coupling \citep{Zhang:2019xya}, in which all $2\rightarrow2$ processes
are considered and one can see clearly how the macroscopic vorticity
in the fluid induces the microscopic polarization of a particle. 

In order to carefully understand the polarization due to spin-orbit
coupling, in this article we consider a simple model to calculate
the polarization of a Dirac fermion scattered by four types of static
potential. We choose the wavepacket as the initial state of the electron
which is the same as \citep{Zhang:2019xya} and sum over the initial
polarization states, so that the initial total angular momentum is
just the OAM of the incident electron. The final state of the electron
is chosen as the common eigenstate of momentum and spin, then the
scattering probability of different final spin can be calculated. 

This article is organised as follows. In Sec. \ref{sec:wavepacket},
we set up the theoretical formulism for the scattering probability
of different final spin. In Sec. \ref{sec:static}, the polarization
of the scattered electron by four types of static potential is studied.
We briefly summarise this article in Sec. \ref{sec:Summary}. 

In this article, the natural unit where $\hbar=c=1$ is adopted. We
choose the metric tensor as $g^{\mu\nu}=\text{diag\,}(+1,-1,-1,-1)$.
Greek indices, such as $\mu,\nu,\rho,\sigma$, run over $0,1,2,3$,
or $t,x,y,z$, while Roman indices, such as $i,j,k$, run over $1,2,3$
or $x,y,z$. The Heaviside-Lorentz convention is chosen for electromagnetism
which is consistent with Peskin and Schroeder \citep{Peskin:1995}.

\section{A wavepacket is scattered by a static potential}

\label{sec:wavepacket}In this section we consider an electron wavepacket
which is scattered by a classical static potential (time independent).
The classical static potential is denoted as $\mathcal{A}$, the center
of which is located at the origin. A very convenient choice for the
static potential is the screened potential model \citep{Gyulassy:1993hr},
whose explicit form in position space is $V(\boldsymbol{x})=Qe^{-d|\boldsymbol{x}|}/(4\pi|\boldsymbol{x}|)$
with electric charge source $Q$ and force distance $1/d$. The form
of this screened potential in momentum space is $V(\boldsymbol{q})=Q/(\boldsymbol{q}^{2}+d^{2})$.
The electron wavepacket is denoted as $\mathcal{B}$. The in-state
$|\phi_{\mathcal{B}},\lambda_{\mathcal{B}};\boldsymbol{b}\rangle_{\text{in}}$
of the single particle wavepacket for this electron in the remote
past with impact parameter $\boldsymbol{b}=(b,0,0)$ can be represented
as 
\begin{equation}
|\phi_{\mathcal{B}},\lambda_{\mathcal{B}};\boldsymbol{b}\rangle_{\text{in}}=\int\frac{d^{3}k}{(2\pi)^{3}}\frac{1}{\sqrt{2E_{k}}}\phi(\boldsymbol{k})e^{-i\boldsymbol{k}\cdot\boldsymbol{b}}|\boldsymbol{k},\lambda_{\mathcal{B}}\rangle_{\text{in}},\label{eq:qq1}
\end{equation}
where $|\boldsymbol{k},\lambda_{\mathcal{B}}\rangle_{\text{in}}$
represents the in-state of a plane wave state of an electron with
momentum $\boldsymbol{k}$ and spin projection component $\lambda_{\mathcal{B}}$
along the direction of positive $y$-axis, and $\phi(\boldsymbol{k})$
is a normalised Guassian wavepacket with momentum center $\boldsymbol{p}_{i}=(0,0,c)$
and momentum width $a$ whose explicit form is
\begin{equation}
\phi(\boldsymbol{k})=\bigg(\frac{(8\pi)^{3}}{a^{6}}\bigg)^{\frac{1}{4}}\exp\bigg(-\frac{(\boldsymbol{k}-\boldsymbol{p}_{i})^{2}}{a^{2}}\bigg).\label{eq:qq2}
\end{equation}
Taking use of $_{\text{in}}\langle\boldsymbol{k}^{\prime},\lambda_{\mathcal{B}}|\boldsymbol{k},\lambda_{\mathcal{B}}\rangle_{\text{in}}=2E_{k}(2\pi)^{3}\delta^{(3)}(\boldsymbol{k}^{\prime}-\boldsymbol{k})$,
we can see that the in-state in Eq. (\ref{eq:qq1}) can be normalised
to $1$ as
\begin{equation}
_{\text{in}}\langle\phi_{\mathcal{B}},\lambda_{\mathcal{B}};\boldsymbol{b}|\phi_{\mathcal{B}},\lambda_{\mathcal{B}};\boldsymbol{b}\rangle_{\text{in}}=\int\frac{d^{3}k}{(2\pi)^{3}}|\phi(\boldsymbol{k})|^{2}=1.\label{eq:qq4}
\end{equation}

Since the electron moves from $-z$-axis to $+z$-axis, we set $c>0$.
We also set $b>0$, then the initial orbital angular momentum of the
electron is along $-y$-axis. When this electron moves toward the
static potential, it will be scattered by some probability. The sacttering
probability with spin component $\lambda_{\mathcal{B}}$ summed over
in the initial state and with spin projection component $\lambda$
along $+y$-axis in the final state is
\begin{eqnarray}
\mathcal{P}(\lambda,b) & = & \frac{1}{2}\int\frac{d^{3}p}{(2\pi)^{3}2E_{p}}\sum_{\lambda_{\mathcal{B}}}\bigg|{}_{\text{out}}\langle\boldsymbol{p},\lambda|\phi_{\mathcal{B}},\lambda_{\mathcal{B}};\boldsymbol{b}\rangle_{\text{in}}\bigg|^{2}\nonumber \\
 & = & \frac{1}{2}\int\frac{d^{3}p}{(2\pi)^{3}2E_{p}}\sum_{\lambda_{\mathcal{B}}}\bigg|\langle\boldsymbol{p},\lambda|\mathcal{S}|\phi_{\mathcal{B}},\lambda_{\mathcal{B}};\boldsymbol{b}\rangle\bigg|^{2}\nonumber \\
 & = & \frac{1}{2}\int\frac{d^{3}p}{(2\pi)^{3}2E_{p}}\sum_{\lambda_{\mathcal{B}}}\bigg|\int\frac{d^{3}k}{(2\pi)^{3}}\frac{1}{\sqrt{2E_{k}}}\phi(\boldsymbol{k})e^{-i\boldsymbol{k}\cdot\boldsymbol{b}}\langle\boldsymbol{p},\lambda|\mathcal{S}|\boldsymbol{k},\lambda_{\mathcal{B}}\rangle\bigg|^{2}\label{eq:qq6}
\end{eqnarray}
where $\mathcal{S}$ denotes the $S$-matrix for the interaction and
we choose the plane wave state $|\boldsymbol{p},\lambda\rangle$ as
the final state with normalization: $\langle\boldsymbol{p}^{\prime},\lambda|\boldsymbol{p},\lambda\rangle=2E_{p}(2\pi)^{3}\delta^{(3)}(\boldsymbol{p}^{\prime}-\boldsymbol{p})$.
Then we can obtain the polarization $\chi(b)$ of the final electron
along $+y$-axis as

\begin{equation}
\chi(b)=\frac{\mathcal{P}(+,b)-\mathcal{P}(-,b)}{\mathcal{P}(+,b)+\mathcal{P}(-,b)}.\label{eq:qq4-1}
\end{equation}

The denominator $\mathcal{P}(+,b)+\mathcal{P}(-,b)$ in Eq. (\ref{eq:qq4-1})
is the normalization factor. In fact, due to the unitarity of the
$S$-matrix, $\mathcal{S}^{\dagger}\mathcal{S}=1$, the sum $\sum_{\lambda}\mathcal{P}(\lambda,b)$
automatically equal to one, i.e.
\begin{eqnarray}
\sum_{\lambda}\mathcal{P}(\lambda,b) & = & \frac{1}{2}\int\frac{d^{3}p}{(2\pi)^{3}2E_{p}}\sum_{\lambda}\sum_{\lambda_{\mathcal{B}}}\bigg|\langle\boldsymbol{p},\lambda|\mathcal{S}|\phi_{\mathcal{B}},\lambda_{\mathcal{B}};\boldsymbol{b}\rangle\bigg|^{2}\nonumber \\
 & = & \frac{1}{2}\int\frac{d^{3}p}{(2\pi)^{3}2E_{p}}\sum_{\lambda}\sum_{\lambda_{\mathcal{B}}}\langle\phi_{\mathcal{B}},\lambda_{\mathcal{B}};\boldsymbol{b}|\mathcal{S}^{\dagger}|\boldsymbol{p},\lambda\rangle\langle\boldsymbol{p},\lambda|\mathcal{S}|\phi_{\mathcal{B}},\lambda_{\mathcal{B}};\boldsymbol{b}\rangle\nonumber \\
 & = & \frac{1}{2}\sum_{\lambda_{\mathcal{B}}}\langle\phi_{\mathcal{B}},\lambda_{\mathcal{B}};\boldsymbol{b}|\mathcal{S}^{\dagger}\mathcal{S}|\phi_{\mathcal{B}},\lambda_{\mathcal{B}};\boldsymbol{b}\rangle\nonumber \\
 & = & \frac{1}{2}\sum_{\lambda_{\mathcal{B}}}\langle\phi_{\mathcal{B}},\lambda_{\mathcal{B}};\boldsymbol{b}|\phi_{\mathcal{B}},\lambda_{\mathcal{B}};\boldsymbol{b}\rangle\nonumber \\
 & = & 1.\label{eq:qq5}
\end{eqnarray}

But in actual calculation, we only consider the process of tree level
and ignore the high order processes, so the sum $\sum_{\lambda}\mathcal{P}(\lambda,b)$
no longer equal to one and the normalization factor in the denominator
in Eq. (\ref{eq:qq4-1}) is necessary.

\section{Polarization of the electron scattered by different static potentials}

\label{sec:static}In this section, we will calculate the polarization
of an electron scattered by different static potentials. The electron
is described by Dirac field. Firstly, we consider the coupling between
Dirac field and static vector field, i.e. the interaction part of
the Lagrangian is $\mathcal{L}_{\text{int}}=-e\bar{\psi}\gamma^{\mu}\psi A_{\mu}$
with $A^{\mu}(t,\boldsymbol{x})=(V(\boldsymbol{x}),\boldsymbol{0})$.
Ignoring the identity operator in $S$-matrix, the term $\langle\boldsymbol{p},\lambda|\mathcal{S}|\boldsymbol{k},\lambda_{\mathcal{B}}\rangle$
in Eq. (\ref{eq:qq6}) at first order in coupling $e$ is
\begin{eqnarray}
\langle\boldsymbol{p},\lambda|\mathcal{S}|\boldsymbol{k},\lambda_{\mathcal{B}}\rangle & = & \langle\boldsymbol{p},\lambda|(-ie)\int d^{4}x\bar{\psi}(x)\gamma^{\mu}\psi(x)A_{\mu}(x)|\boldsymbol{k},\lambda_{\mathcal{B}}\rangle\nonumber \\
 & \equiv & (-ie)\bar{u}(p,\lambda)\gamma\cdot A(\boldsymbol{p}-\boldsymbol{k})u(k,\lambda_{\mathcal{B}})(2\pi)\delta(E_{p}-E_{k}),\label{eq:qq7}
\end{eqnarray}
where $A^{\mu}(\boldsymbol{p}-\boldsymbol{k})=(V(\boldsymbol{p}-\boldsymbol{k}),\boldsymbol{0})$
and $u(k,\lambda_{\mathcal{B}}),u(p,\lambda)$ are solved in chiral
representation of Dirac matrix for free Dirac equation. In the following
calculation, we will choose $u(k,\lambda_{\mathcal{B}}),u(p,\lambda)$
as
\begin{eqnarray}
u(k,\lambda_{\mathcal{B}}) & = & \frac{1}{\sqrt{2(m+E_{k})}}\bigg(\begin{array}{c}
(m+E_{k}-\boldsymbol{\sigma}\cdot\boldsymbol{k})\xi_{\lambda_{\mathcal{B}}}\\
(m+E_{k}+\boldsymbol{\sigma}\cdot\boldsymbol{k})\xi_{\lambda_{\mathcal{B}}}
\end{array}\bigg)\nonumber \\
u(p,\lambda) & = & \frac{1}{\sqrt{2(m+E_{p})}}\bigg(\begin{array}{c}
(m+E_{p}-\boldsymbol{\sigma}\cdot\boldsymbol{p})\xi_{\lambda}\\
(m+E_{p}+\boldsymbol{\sigma}\cdot\boldsymbol{p})\xi_{\lambda}
\end{array}\bigg),\label{eq:qq8}
\end{eqnarray}
where $\xi_{\lambda}$ is the eigenstate of spin operator $\boldsymbol{\sigma}$
along $+y$-axis with eigenvalue $\lambda$, i.e. $\xi_{\lambda}^{\dagger}\boldsymbol{\sigma}\xi_{\lambda}=\lambda\hat{\boldsymbol{y}}$
with $\lambda=\pm1$. Since $\lambda_{\mathcal{B}}$ will be summed
over later, the explicit form of $\xi_{\lambda_{\mathcal{B}}}$ is
irrelevant. 

From Eq. (\ref{eq:qq6}), we can obtain the scattering probability
$\mathcal{P}^{V}(\lambda,b)$ with the final spin projection component
$\lambda$ and collision parameter $b$ for static vector potential,
which is calculated in detail in Appendix \ref{sec:calculation for P}.
We list the explicit form of $\mathcal{P}^{V}(\lambda,b)$ as follows,
\begin{equation}
\mathcal{P}^{V}(\lambda,b)=\mathcal{P}_{0}^{V}(b)+\lambda\mathcal{P}_{1}^{V}(b),\label{eq:qq9}
\end{equation}
with unpolarised probability $\mathcal{P}_{0}^{V}(b)$ and polarised
probability $\mathcal{P}_{1}^{V}(b)$ defined as 7-dimensional integrals,
\begin{eqnarray*}
\mathcal{P}_{0}^{V}(b) & = & \frac{\alpha}{4(2\pi)^{6}}\int dpp^{4}\int d\Omega_{p}\int d\Omega d\Omega^{\prime}A_{0}(\boldsymbol{p}-\boldsymbol{k})A_{0}(\boldsymbol{p}-\boldsymbol{k}^{\prime})\phi(\boldsymbol{k})\phi(\boldsymbol{k}^{\prime})\cos[pb\hat{\boldsymbol{x}}\cdot(\hat{\boldsymbol{k}}-\hat{\boldsymbol{k}}^{\prime})]\\
 &  & \times[(E_{p}+m)^{2}+(E_{p}-m)^{2}\hat{\boldsymbol{k}}\cdot\hat{\boldsymbol{k}}^{\prime}+p^{2}\hat{\boldsymbol{p}}\cdot(\hat{\boldsymbol{k}}+\hat{\boldsymbol{k}}^{\prime})]
\end{eqnarray*}
\begin{eqnarray*}
\mathcal{P}_{1}^{V}(b) & = & \frac{\alpha}{4(2\pi)^{6}}\int dpp^{4}\int d\Omega_{p}\int d\Omega d\Omega^{\prime}A_{0}(\boldsymbol{p}-\boldsymbol{k})A_{0}(\boldsymbol{p}-\boldsymbol{k}^{\prime})\phi(\boldsymbol{k})\phi(\boldsymbol{k}^{\prime})\sin[pb\hat{\boldsymbol{x}}\cdot(\hat{\boldsymbol{k}}-\hat{\boldsymbol{k}}^{\prime})]\\
 &  & \times\bigg((E_{p}-m)^{2}(2\hat{\boldsymbol{y}}\cdot\hat{\boldsymbol{p}}\hat{\boldsymbol{p}}-\hat{\boldsymbol{y}})\cdot(\hat{\boldsymbol{k}}\times\hat{\boldsymbol{k}}^{\prime})+p^{2}\hat{\boldsymbol{y}}\cdot[\hat{\boldsymbol{p}}\times(\hat{\boldsymbol{k}}-\hat{\boldsymbol{k}}^{\prime})]\bigg)
\end{eqnarray*}
where $d\Omega_{p}$, $d\Omega$, $d\Omega^{\prime}$ represent the
differential solid angle of $\hat{\boldsymbol{p}}$, $\hat{\boldsymbol{k}}$,
$\hat{\boldsymbol{k}}^{\prime}$ respectively, and $\hat{\boldsymbol{x}}=(1,0,0)$,
$\hat{\boldsymbol{y}}=(0,1,0)$.

Then the polarization $\chi^{V}(b)$ of the electron scattered by
the static vector potential in Eq. (\ref{eq:qq4-1}) becomes
\begin{equation}
\chi^{V}(b)=\frac{\mathcal{P}_{1}^{V}(b)}{\mathcal{P}_{0}^{V}(b)}.\label{eq:qq11}
\end{equation}

If the electron is scattered by the static pseudovector, scalar and
pseudoscalar potentials respectively, then the interaction part of
the Lagrangian $\mathcal{L}_{\text{int}}$ becomes $\bar{\psi}\gamma^{\mu}\gamma^{5}\psi A_{\mu},\bar{\psi}\psi\phi,\bar{\psi}\gamma^{5}\psi\phi$,
where $\phi$ is a scalar or pseudoscalar field. In Appendix \ref{sec:calculation for P},
we also calculate the polarization function $\chi(b)$ for these three
types of static potential. 

\begin{figure}[H]
\begin{centering}
\includegraphics[scale=0.35]{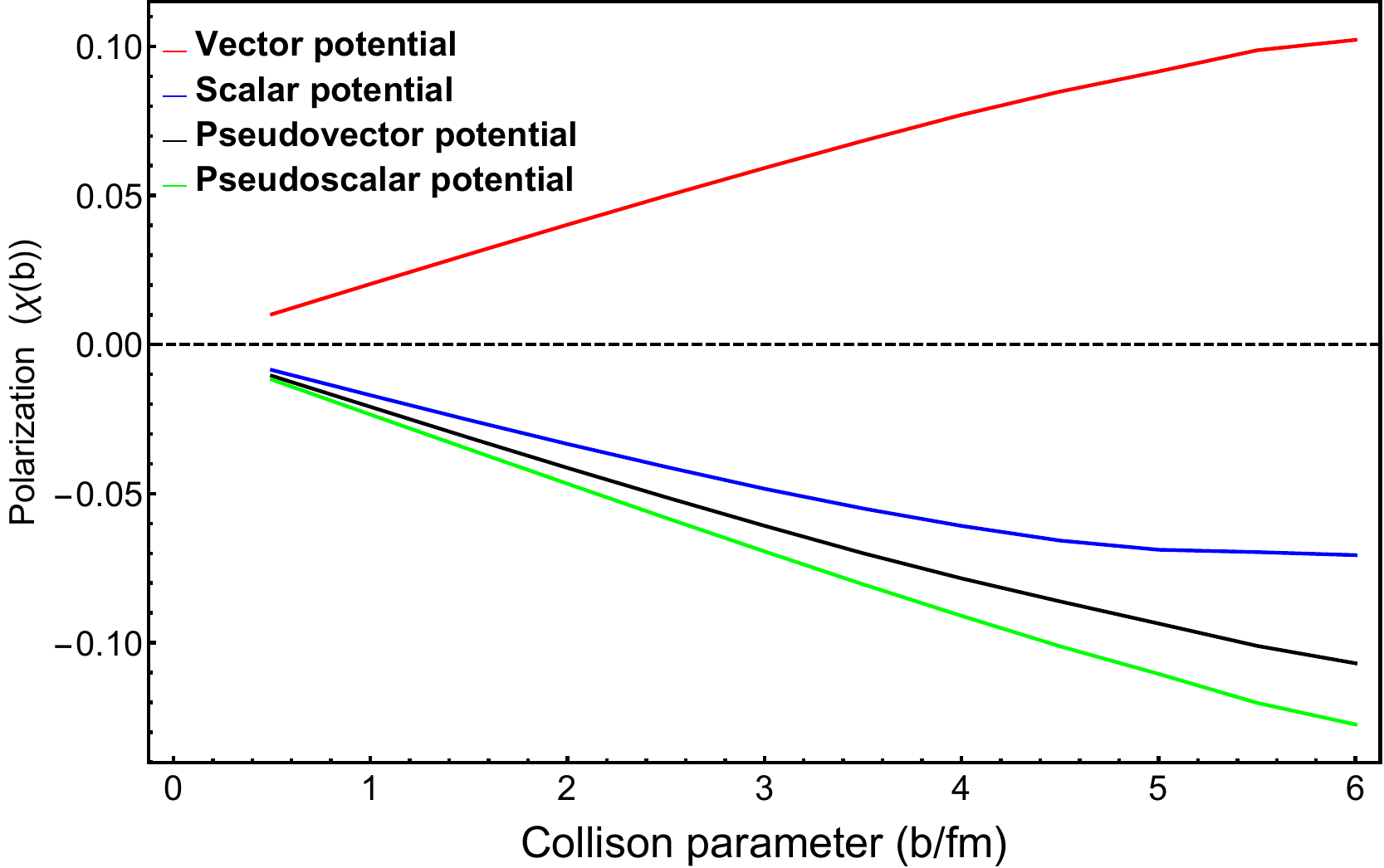}
\par\end{centering}
\caption{\label{fig:1}Polarization $\chi(b)$ as a fucntion of collision parameter
$b$}

\end{figure}

Figure \ref{fig:1} shows the numerical results of the polarization
$\chi(b)$ as a fucntion of collision parameter $b$ for four types
of static potential with the parameters, where we take following parameters:
initial energy of the incident electron $c=1\,\mathrm{GeV}$, the
width of wavepacket $a=0.1\,\mathrm{GeV},$Debye screen mass $d=0.1\,\mathrm{GeV}$,
and electron mass $m=0.000511\,\mathrm{GeV}$. Since $b>0$, $c>0$,
i.e. the initial OAM of the electron is along negative $y$-axis,
we may expect that the final eletron scattered by the static potentials
polarises more preferentially along negative $y$-axis, which is consistent
with the results for pseudovector, scalar and pseudoscalar potentials
as shown in Figure \ref{fig:1}. However, for vector potential, the
final eletron polarises more preferentially along positive $y$-axis,
which is opposite to the direction of the initial OAM of the electron.
This inconsistence may result from the fact that the virtual photon
exchanged by the electron and the static potential also carries spin
angular momentum of $1\,\hbar$, leading to the opposite polarization
of the final electron due to the conservation of angular momentum.
In Figure \ref{fig:1}, we only plot the curves in the range $0<b<6\,\mathrm{fm}$,
where the the absolute value of polarization becomes larger as $b$
increases. Especially in the range $0<b<2\,\mathrm{fm}$, the magnitude
order of the polarization value is the same as the recent experimental
result \citep{STAR:2017ckg}. For the range $b>6\,\mathrm{fm}$, the
numerical result becomes unstable, which is not shown in the plot.
It is expected that the polarization $\chi(b)$ becomes zero as $b$
tends to be very large, since in this case the influence of the static
potential on the electron is very weak.

\section{Summary}

\label{sec:Summary}In this article, we calculate the polarization
of an electron scattered by four different types of static potential.
Through the scattering of static potentials, it is expected that the
initial OAM of the incident electron can transfer into the final spin
angular momentum, which is consistent with the numerical results for
pseudovector, scalar and pseudoscalar potentials. However the electron
polarises more preferentially opposite to the initial OAM of the electron,
which may result from the spin of the virtual photon exchanged. For
the range of the collision parameter $0<b<2\,\mathrm{fm}$, the magnitude
order of the polarization value is consistent with the recent experimental
result.

\section{\textup{Acknowledgments}}

We thank Qun Wang for helpful discussion. This work was supported
in part by the Science and Technology Research Project of Education
Department of Hubei Province under Grant No. D20221901, and the National
Natural Science Foundation of China (NSFC) under Grant Nos. 12265013
and 12073008.

\appendix

\section{Calculation for $\mathcal{P}(\lambda,b)$}

\label{sec:calculation for P}From Eq. (\ref{eq:qq7}) and Eq. (\ref{eq:qq1}),
we obtain

\begin{eqnarray}
 &  & \langle\boldsymbol{p},\lambda|\mathcal{S}|\phi_{\mathcal{B}},\lambda_{\mathcal{B}}\rangle\nonumber \\
 & = & \int\frac{d^{3}k}{(2\pi)^{3}}\frac{\phi(\boldsymbol{k})}{\sqrt{2E_{k}}}e^{-i\boldsymbol{k}\cdot\boldsymbol{b}}(-ie)\bar{u}(p,\lambda)\gamma\cdot A(\boldsymbol{p}-\boldsymbol{k})u(k,\lambda_{\mathcal{B}})(2\pi)\delta(E_{p}-E_{k})\nonumber \\
 & = & \frac{1}{(2\pi)^{2}}\int_{0}^{\infty}d|\boldsymbol{k}|\boldsymbol{k}^{2}\int d\Omega\frac{\phi(\boldsymbol{k})}{\sqrt{2E_{k}}}e^{-i\boldsymbol{k}\cdot\boldsymbol{b}}(-ie)\bar{u}(p,\lambda)\gamma\cdot A(\boldsymbol{p}-\boldsymbol{k})u(k,\lambda_{\mathcal{B}})\frac{E_{p}}{|\boldsymbol{p}|}\delta(|\boldsymbol{k}|-|\boldsymbol{p}|)\nonumber \\
 & = & \frac{-ie|\boldsymbol{p}|}{(2\pi)^{2}}\sqrt{\frac{E_{p}}{2}}\int d\Omega A_{0}(\boldsymbol{p}-\boldsymbol{k})\phi(\boldsymbol{k})e^{-i\boldsymbol{k}\cdot\boldsymbol{b}}u^{\dagger}(p,\lambda)u(k,\lambda_{\mathcal{B}}),\label{eq:qq10}
\end{eqnarray}

where $d\Omega$ represents the solid angle of $\hat{\boldsymbol{k}}$.
Taking square of Eq. (\ref{eq:qq10}) and summing over $\lambda_{\mathcal{B}}$
gives

\begin{eqnarray}
 &  & \sum_{\lambda_{\mathcal{B}}}\bigg|\langle\boldsymbol{p},\lambda|\mathcal{S}|\phi_{\mathcal{B}},\lambda_{\mathcal{B}}\rangle\bigg|^{2}\nonumber \\
 & = & \frac{\alpha\boldsymbol{p}^{2}E_{p}}{(2\pi)^{3}}\int d\Omega d\Omega^{\prime}A_{0}(\boldsymbol{p}-\boldsymbol{k})A_{0}(\boldsymbol{p}-\boldsymbol{k}^{\prime})\phi(\boldsymbol{k})\phi(\boldsymbol{k}^{\prime})e^{-i(\boldsymbol{k}-\boldsymbol{k}^{\prime})\cdot\boldsymbol{b}}\nonumber \\
 &  & \times\sum_{\lambda_{\mathcal{B}}}[u^{\dagger}(p,\lambda)u(k,\lambda_{\mathcal{B}})u^{\dagger}(k^{\prime},\lambda_{\mathcal{B}})u(p,\lambda)]\label{eq:qq11-1}
\end{eqnarray}

where $d\Omega^{\prime}$ represents the solid angle of $\hat{\boldsymbol{k}}^{\prime}$.
In the following, we will calculate the third line of Eq. (\ref{eq:qq11-1})
in detail. We can see
\begin{equation}
\sum_{\lambda_{\mathcal{B}}}[u^{\dagger}(p,\lambda)u(k,\lambda_{\mathcal{B}})u^{\dagger}(k^{\prime},\lambda_{\mathcal{B}})u(p,\lambda)]=\sum_{\lambda_{\mathcal{B}}}\text{tr}[u(p,\lambda)u^{\dagger}(p,\lambda)u(k,\lambda_{\mathcal{B}})u^{\dagger}(k^{\prime},\lambda_{\mathcal{B}})]\label{eq:qq12}
\end{equation}

For $u(k,\lambda_{\mathcal{B}})u^{\dagger}(k^{\prime},\lambda_{\mathcal{B}})$
part, we have

\begin{eqnarray*}
\text{M}_{\mathcal{B}} & = & \sum_{\lambda_{\mathcal{B}}}u(k,\lambda_{\mathcal{B}})u^{\dagger}(k^{\prime},\lambda_{\mathcal{B}})\\
 & = & \frac{1}{2\sqrt{(m+E_{k})(m+E_{k^{\prime}})}}\\
 &  & \times\sum_{\lambda_{\mathcal{B}}}\bigg(\begin{array}{c}
(m+E_{k}-\boldsymbol{\sigma}\cdot\boldsymbol{k})\xi_{\mathcal{B}}\\
(m+E_{k}+\boldsymbol{\sigma}\cdot\boldsymbol{k})\xi_{\mathcal{B}}
\end{array}\bigg)\left(\begin{array}{cc}
\xi_{\mathcal{B}}^{\dagger}(m+E_{k^{\prime}}-\boldsymbol{\sigma}\cdot\boldsymbol{k}^{\prime}),\xi_{\mathcal{B}}^{\dagger}( & m+E_{k^{\prime}}+\boldsymbol{\sigma}\cdot\boldsymbol{k}^{\prime})\end{array}\right)\\
 & = & \frac{1}{2\sqrt{(m+E_{k})(m+E_{k^{\prime}})}}\\
 &  & \times\left(\begin{array}{cc}
(m+E_{k}-\boldsymbol{\sigma}\cdot\boldsymbol{k})(m+E_{k^{\prime}}-\boldsymbol{\sigma}\cdot\boldsymbol{k}^{\prime}), & (m+E_{k}-\boldsymbol{\sigma}\cdot\boldsymbol{k})(m+E_{k^{\prime}}+\boldsymbol{\sigma}\cdot\boldsymbol{k}^{\prime})\\
(m+E_{k}+\boldsymbol{\sigma}\cdot\boldsymbol{k})(m+E_{k^{\prime}}-\boldsymbol{\sigma}\cdot\boldsymbol{k}^{\prime}), & (m+E_{k}+\boldsymbol{\sigma}\cdot\boldsymbol{k})(m+E_{k^{\prime}}+\boldsymbol{\sigma}\cdot\boldsymbol{k}^{\prime})
\end{array}\right)
\end{eqnarray*}

For $u(p,\lambda)u^{\dagger}(p,\lambda)$ part, we have
\begin{equation}
u(p,\lambda)u^{\dagger}(p,\lambda)=u(p,\lambda)\bar{u}(p,\lambda)\gamma^{0}=\frac{1}{2}(1+\lambda\gamma^{5}\gamma\cdot S)(\gamma\cdot p+m)\gamma^{0}\label{eq:q14}
\end{equation}

where $S^{\mu}$ is the 4-dimensional spin vector

\begin{equation}
S^{\mu}=(S^{0},\boldsymbol{S})=\bigg(\frac{\boldsymbol{p}\cdot\hat{\boldsymbol{y}}}{m},\hat{\boldsymbol{y}}+\frac{(\hat{\boldsymbol{y}}\cdot\boldsymbol{p})\boldsymbol{p}}{m(m+E_{p})}\bigg)\label{eq:q15}
\end{equation}

We can see
\[
\gamma\cdot p+m=\left(\begin{array}{cc}
m & E_{p}-\boldsymbol{\sigma}\cdot\boldsymbol{p}\\
E_{p}+\boldsymbol{\sigma}\cdot\boldsymbol{p} & m
\end{array}\right)
\]
\[
\gamma^{5}\gamma\cdot S=\left(\begin{array}{cc}
0 & -S^{0}+\boldsymbol{\sigma}\cdot\boldsymbol{S}\\
S^{0}+\boldsymbol{\sigma}\cdot\boldsymbol{S} & 0
\end{array}\right)
\]

The term related to $\lambda$ in $u(p,\lambda)u^{\dagger}(p,\lambda)$
is
\begin{eqnarray*}
\text{M}_{1} & = & \frac{\lambda}{2}\gamma^{5}\gamma\cdot S(\gamma\cdot p+m)\gamma^{0}\\
 & = & \frac{\lambda}{2}\left(\begin{array}{cc}
0 & -S^{0}+\boldsymbol{\sigma}\cdot\boldsymbol{S}\\
S^{0}+\boldsymbol{\sigma}\cdot\boldsymbol{S} & 0
\end{array}\right)\left(\begin{array}{cc}
m & E_{p}-\boldsymbol{\sigma}\cdot\boldsymbol{p}\\
E_{p}+\boldsymbol{\sigma}\cdot\boldsymbol{p} & m
\end{array}\right)\left(\begin{array}{cc}
0 & 1\\
1 & 0
\end{array}\right)\\
 & = & \frac{\lambda}{2}\left(\begin{array}{cc}
(-S^{0}+\boldsymbol{\sigma}\cdot\boldsymbol{S})m & (-S^{0}+\boldsymbol{\sigma}\cdot\boldsymbol{S})(E_{p}+\boldsymbol{\sigma}\cdot\boldsymbol{p})\\
(S^{0}+\boldsymbol{\sigma}\cdot\boldsymbol{S})(E_{p}-\boldsymbol{\sigma}\cdot\boldsymbol{p}) & (S^{0}+\boldsymbol{\sigma}\cdot\boldsymbol{S})m
\end{array}\right)
\end{eqnarray*}

The term without $\lambda$ in $u(p,\lambda)u^{\dagger}(p,\lambda)$
is
\[
\text{M}_{0}=\frac{1}{2}(\gamma\cdot p+m)\gamma^{0}=\frac{1}{2}\left(\begin{array}{cc}
m & E_{p}-\boldsymbol{\sigma}\cdot\boldsymbol{p}\\
E_{p}+\boldsymbol{\sigma}\cdot\boldsymbol{p} & m
\end{array}\right)\left(\begin{array}{cc}
0 & 1\\
1 & 0
\end{array}\right)=\frac{1}{2}\left(\begin{array}{cc}
E_{p}-\boldsymbol{\sigma}\cdot\boldsymbol{p} & m\\
m & E_{p}+\boldsymbol{\sigma}\cdot\boldsymbol{p}
\end{array}\right)
\]

The term related to $\lambda$ in $\sum_{\lambda_{\mathcal{B}}}[u^{\dagger}(p,\lambda)u(k,\lambda_{\mathcal{B}})u^{\dagger}(k^{\prime},\lambda_{\mathcal{B}})u(p,\lambda)]$
is
\begin{eqnarray*}
\text{tr}\,(\text{M}_{1}\text{M}_{\mathcal{B}}) & = & (\text{M}_{1}\text{M}_{\mathcal{B}})_{11}+(\text{M}_{1}\text{M}_{\mathcal{B}})_{22}\\
 & = & (\text{M}_{1})_{11}(\text{M}_{\mathcal{B}})_{11}+(\text{M}_{1})_{12}(\text{M}_{\mathcal{B}})_{21}+(\text{M}_{1})_{21}(\text{M}_{\mathcal{B}})_{12}+(\text{M}_{1})_{22}(\text{M}_{\mathcal{B}})_{22}\\
 & \equiv & \frac{\lambda}{\sqrt{(m+E_{k})(m+E_{k^{\prime}})}}\times\frac{1}{4}(\text{I}_{1}+\text{II}_{1}+\text{III}_{1}+\text{IV}_{1})
\end{eqnarray*}

where
\begin{eqnarray*}
\text{I}_{1} & = & \text{tr}\,(-S^{0}+\boldsymbol{\sigma}\cdot\boldsymbol{S})m(m+E_{k}-\boldsymbol{\sigma}\cdot\boldsymbol{k})(m+E_{k^{\prime}}-\boldsymbol{\sigma}\cdot\boldsymbol{k}^{\prime})\\
\text{II}_{1} & = & \text{tr}\,(-S^{0}+\boldsymbol{\sigma}\cdot\boldsymbol{S})(E_{p}+\boldsymbol{\sigma}\cdot\boldsymbol{p})(m+E_{k}+\boldsymbol{\sigma}\cdot\boldsymbol{k})(m+E_{k^{\prime}}-\boldsymbol{\sigma}\cdot\boldsymbol{k}^{\prime})\\
\text{III}_{1} & = & \text{tr}\,(S^{0}+\boldsymbol{\sigma}\cdot\boldsymbol{S})(E_{p}-\boldsymbol{\sigma}\cdot\boldsymbol{p})(m+E_{k}-\boldsymbol{\sigma}\cdot\boldsymbol{k})(m+E_{k^{\prime}}+\boldsymbol{\sigma}\cdot\boldsymbol{k}^{\prime})\\
\text{IV}_{1} & = & \text{tr}\,(S^{0}+\boldsymbol{\sigma}\cdot\boldsymbol{S})m(m+E_{k}+\boldsymbol{\sigma}\cdot\boldsymbol{k})(m+E_{k^{\prime}}+\boldsymbol{\sigma}\cdot\boldsymbol{k}^{\prime})
\end{eqnarray*}

Taking use of 
\[
\frac{1}{2}\text{tr}\,(\boldsymbol{\sigma}\cdot\boldsymbol{a})(\boldsymbol{\sigma}\cdot\boldsymbol{b})=\boldsymbol{a}\cdot\boldsymbol{b}
\]
\[
\frac{1}{2}\text{tr}\,(\boldsymbol{\sigma}\cdot\boldsymbol{a})(\boldsymbol{\sigma}\cdot\boldsymbol{b})(\boldsymbol{\sigma}\cdot\boldsymbol{c})=i\boldsymbol{a}\cdot(\boldsymbol{b}\times\boldsymbol{c})
\]
\[
\frac{1}{2}\text{tr}\,(\boldsymbol{\sigma}\cdot\boldsymbol{a})(\boldsymbol{\sigma}\cdot\boldsymbol{b})(\boldsymbol{\sigma}\cdot\boldsymbol{c})(\boldsymbol{\sigma}\cdot\boldsymbol{d})=(\boldsymbol{a}\cdot\boldsymbol{b})(\boldsymbol{c}\cdot\boldsymbol{d})-(\boldsymbol{a}\cdot\boldsymbol{c})(\boldsymbol{b}\cdot\boldsymbol{d})+(\boldsymbol{a}\cdot\boldsymbol{d})(\boldsymbol{b}\cdot\boldsymbol{c})
\]

we have
\[
\frac{1}{4}(\text{I}_{1}+\text{IV}_{1})=im\boldsymbol{S}\cdot(\boldsymbol{k}\times\boldsymbol{k}^{\prime})
\]
\[
\frac{1}{4}(\text{II}_{1}+\text{III}_{1})=i(S^{0}\boldsymbol{p}-E_{p}\boldsymbol{S})\cdot(\boldsymbol{k}\times\boldsymbol{k}^{\prime})+i(m+E_{p})\boldsymbol{S}\cdot[\boldsymbol{p}\times(\boldsymbol{k}-\boldsymbol{k}^{\prime})]
\]

Note that $E_{k}=E_{k^{\prime}}=E_{p}$ due to the delta function
$\delta(E_{p}-E_{k})$ in Eq. (\ref{eq:qq7}). Finally we have
\begin{eqnarray*}
\text{tr}\,(\text{M}_{1}\text{M}_{\mathcal{B}}) & = & \frac{\lambda}{\sqrt{(m+E_{k})(m+E_{k^{\prime}})}}\times\frac{1}{4}(\text{I}_{1}+\text{II}_{1}+\text{III}_{1}+\text{IV}_{1})\\
 & = & i\lambda\bigg((E_{p}-m)^{2}(2\hat{\boldsymbol{y}}\cdot\hat{\boldsymbol{p}}\hat{\boldsymbol{p}}-\hat{\boldsymbol{y}})\cdot(\hat{\boldsymbol{k}}\times\hat{\boldsymbol{k}}^{\prime})+p^{2}\hat{\boldsymbol{y}}\cdot[\hat{\boldsymbol{p}}\times(\hat{\boldsymbol{k}}-\hat{\boldsymbol{k}}^{\prime})]\bigg)
\end{eqnarray*}

Then we can obtain the coefficient of $\lambda$ in Eq. (\ref{eq:qq6})
as
\begin{eqnarray*}
\mathcal{P}_{1}(b) & = & \frac{\alpha}{4(2\pi)^{6}}\int dpp^{4}\int d\Omega_{p}\int d\Omega d\Omega^{\prime}A_{0}(\boldsymbol{p}-\boldsymbol{k})A_{0}(\boldsymbol{p}-\boldsymbol{k}^{\prime})\phi(\boldsymbol{k})\phi(\boldsymbol{k}^{\prime})\sin[pb\hat{\boldsymbol{x}}\cdot(\hat{\boldsymbol{k}}-\hat{\boldsymbol{k}}^{\prime})]\\
 &  & \times\bigg((E_{p}-m)^{2}(2\hat{\boldsymbol{y}}\cdot\hat{\boldsymbol{p}}\hat{\boldsymbol{p}}-\hat{\boldsymbol{y}})\cdot(\hat{\boldsymbol{k}}\times\hat{\boldsymbol{k}}^{\prime})+p^{2}\hat{\boldsymbol{y}}\cdot[\hat{\boldsymbol{p}}\times(\hat{\boldsymbol{k}}-\hat{\boldsymbol{k}}^{\prime})]\bigg)
\end{eqnarray*}

where $\alpha=e^{2}/4\pi$, $p=|\boldsymbol{p}|=|\boldsymbol{k}|=|\boldsymbol{k}^{\prime}|$
and $d\Omega_{p}$ represents the solid angle of $\hat{\boldsymbol{p}}$.
The term errelevant to $\lambda$ in $\sum_{\lambda_{\mathcal{B}}}[u^{\dagger}(p,\lambda)u(k,\lambda_{\mathcal{B}})u^{\dagger}(k^{\prime},\lambda_{\mathcal{B}})u(p,\lambda)]$
is
\begin{eqnarray*}
\text{tr}\,(\text{M}_{0}\text{M}_{\mathcal{B}}) & = & (\text{M}_{0}\text{M}_{\mathcal{B}})_{11}+(\text{M}_{0}\text{M}_{\mathcal{B}})_{22}\\
 & = & (\text{M}_{0})_{11}(\text{M}_{\mathcal{B}})_{11}+(\text{M}_{0})_{12}(\text{M}_{\mathcal{B}})_{21}+(\text{M}_{0})_{21}(\text{M}_{\mathcal{B}})_{12}+(\text{M}_{0})_{22}(\text{M}_{\mathcal{B}})_{22}\\
 & \equiv & \frac{1}{m+E_{p}}\times\frac{1}{4}(\text{I}_{0}+\text{II}_{0}+\text{III}_{0}+\text{IV}_{0})
\end{eqnarray*}

where
\begin{eqnarray*}
\text{I}_{0} & = & \text{tr}\,(E_{p}-\boldsymbol{\sigma}\cdot\boldsymbol{p})(m+E_{p}-\boldsymbol{\sigma}\cdot\boldsymbol{k})(m+E_{p}-\boldsymbol{\sigma}\cdot\boldsymbol{k}^{\prime})\\
\text{II}_{0} & = & \text{tr}\,m(m+E_{p}+\boldsymbol{\sigma}\cdot\boldsymbol{k})(m+E_{p}-\boldsymbol{\sigma}\cdot\boldsymbol{k}^{\prime})\\
\text{III}_{0} & = & \text{tr}\,m(m+E_{p}-\boldsymbol{\sigma}\cdot\boldsymbol{k})(m+E_{p}+\boldsymbol{\sigma}\cdot\boldsymbol{k}^{\prime})\\
\text{IV}_{0} & = & \text{tr}\,(E_{p}+\boldsymbol{\sigma}\cdot\boldsymbol{p})(m+E_{p}+\boldsymbol{\sigma}\cdot\boldsymbol{k})(m+E_{p}+\boldsymbol{\sigma}\cdot\boldsymbol{k}^{\prime})
\end{eqnarray*}

We have
\[
\frac{1}{4}(\text{I}_{0}+\text{IV}_{0})=E_{p}(m+E_{p})^{2}+E_{p}\boldsymbol{k}\cdot\boldsymbol{k}^{\prime}+(m+E_{p})\boldsymbol{p}\cdot\boldsymbol{k}+(m+E_{p})\boldsymbol{p}\cdot\boldsymbol{k}^{\prime}
\]
\[
\frac{1}{4}(\text{II}_{0}+\text{III}_{0})=m(m+E_{p})^{2}-m\boldsymbol{k}\cdot\boldsymbol{k}^{\prime}
\]

then
\[
\frac{1}{4}(\text{I}_{0}+\text{II}_{0}+\text{III}_{0}+\text{IV}_{0})=(m+E_{p})^{3}+(E_{p}-m)\boldsymbol{k}\cdot\boldsymbol{k}^{\prime}+(m+E_{p})\boldsymbol{p}\cdot(\boldsymbol{k}+\boldsymbol{k}^{\prime})
\]

Finally we have
\begin{eqnarray*}
\text{tr}\,(\text{M}_{0}\text{M}_{\mathcal{B}}) & = & \frac{1}{m+E_{p}}\times\frac{1}{4}(\text{I}_{0}+\text{II}_{0}+\text{III}_{0}+\text{IV}_{0})\\
 & = & (E_{p}+m)^{2}+(E_{p}-m)^{2}\hat{\boldsymbol{k}}\cdot\hat{\boldsymbol{k}}^{\prime}+p^{2}\hat{\boldsymbol{p}}\cdot(\hat{\boldsymbol{k}}+\hat{\boldsymbol{k}}^{\prime})
\end{eqnarray*}

Then we can obtain the term without $\lambda$ in Eq. (\ref{eq:qq6})
as
\begin{eqnarray*}
\mathcal{P}_{0}(b) & = & \frac{\alpha}{4(2\pi)^{6}}\int dpp^{4}\int d\Omega_{p}\int d\Omega d\Omega^{\prime}A_{0}(\boldsymbol{p}-\boldsymbol{k})A_{0}(\boldsymbol{p}-\boldsymbol{k}^{\prime})\phi(\boldsymbol{k})\phi(\boldsymbol{k}^{\prime})\cos[pb\hat{\boldsymbol{x}}\cdot(\hat{\boldsymbol{k}}-\hat{\boldsymbol{k}}^{\prime})]\\
 &  & \times[(E_{p}+m)^{2}+(E_{p}-m)^{2}\hat{\boldsymbol{k}}\cdot\hat{\boldsymbol{k}}^{\prime}+p^{2}\hat{\boldsymbol{p}}\cdot(\hat{\boldsymbol{k}}+\hat{\boldsymbol{k}}^{\prime})]
\end{eqnarray*}

The probability $\mathcal{P}(\lambda,b)$ at impact parameter $b$
becomes
\[
\mathcal{P}(\lambda,b)=\mathcal{P}_{0}(b)+\lambda\mathcal{P}_{1}(b)
\]

For the potentials of pseudovector, scalar and pseudoscalar, we can
replace the $\gamma^{\mu}A_{\mu}$ factor in the interaction part
of the Lagrangian $\mathcal{L}_{\text{int}}=-e\bar{\psi}\gamma^{\mu}\psi A_{\mu}$
by $\gamma^{\mu}\gamma^{5}A_{\mu},\phi,\gamma^{5}\phi$, where $\phi$
is a scalar or pseudoscalar field. The calculations for the scattering
probability by this three types of interaction are similar to the
vector case, and we list the results as follows.

For pseudovector potential, we have
\begin{eqnarray*}
\mathcal{P}_{1}^{PV}(b) & = & \frac{\alpha}{4(2\pi)^{6}}\int dpp^{6}\int d\Omega_{p}\int d\Omega d\Omega^{\prime}A_{0}(\boldsymbol{p}-\boldsymbol{k})A_{0}(\boldsymbol{p}-\boldsymbol{k}^{\prime})\phi(\boldsymbol{k})\phi(\boldsymbol{k}^{\prime})\cos[pb\hat{\boldsymbol{x}}\cdot(\hat{\boldsymbol{k}}-\hat{\boldsymbol{k}}^{\prime})]\\
 &  & \times\{\hat{\boldsymbol{y}}\cdot(\boldsymbol{k}\times\boldsymbol{k}^{\prime})-\hat{\boldsymbol{y}}\cdot[\hat{\boldsymbol{p}}\times(\hat{\boldsymbol{k}}-\hat{\boldsymbol{k}}^{\prime})]\}
\end{eqnarray*}
\begin{eqnarray*}
\mathcal{P}_{0}^{PV}(b) & = & \frac{\alpha}{4(2\pi)^{6}}\int dpp^{6}\int d\Omega_{p}\int d\Omega d\Omega^{\prime}A_{0}(\boldsymbol{p}-\boldsymbol{k})A_{0}(\boldsymbol{p}-\boldsymbol{k}^{\prime})\phi(\boldsymbol{k})\phi(\boldsymbol{k}^{\prime})\sin[pb\hat{\boldsymbol{x}}\cdot(\hat{\boldsymbol{k}}-\hat{\boldsymbol{k}}^{\prime})]\\
 &  & \times(1+\hat{\boldsymbol{k}}\cdot\hat{\boldsymbol{k}}^{\prime}+\hat{\boldsymbol{p}}\cdot\hat{\boldsymbol{k}}^{\prime}+\hat{\boldsymbol{k}}\cdot\hat{\boldsymbol{p}})
\end{eqnarray*}

For scalar potential, we have
\begin{eqnarray*}
\mathcal{P}_{1}^{S}(b) & = & \frac{\alpha}{4(2\pi)^{6}}\int dpp^{4}\int d\Omega_{p}\int d\Omega d\Omega^{\prime}A_{0}(\boldsymbol{p}-\boldsymbol{k})A_{0}(\boldsymbol{p}-\boldsymbol{k}^{\prime})\phi(\boldsymbol{k})\phi(\boldsymbol{k}^{\prime})\sin[pb\hat{\boldsymbol{x}}\cdot(\hat{\boldsymbol{k}}-\hat{\boldsymbol{k}}^{\prime})]\\
 &  & \times\{(E_{p}-m)^{2}(2\hat{\boldsymbol{y}}\cdot\hat{\boldsymbol{p}}\hat{\boldsymbol{p}}-\hat{\boldsymbol{y}})\cdot(\hat{\boldsymbol{k}}\times\hat{\boldsymbol{k}}^{\prime})-p^{2}\hat{\boldsymbol{y}}\cdot[\hat{\boldsymbol{p}}\times(\hat{\boldsymbol{k}}-\hat{\boldsymbol{k}}^{\prime})]\}
\end{eqnarray*}
\begin{eqnarray*}
\mathcal{P}_{0}^{S}(b) & = & \frac{\alpha}{4(2\pi)^{6}}\int dpp^{4}\int d\Omega_{p}\int d\Omega d\Omega^{\prime}A_{0}(\boldsymbol{p}-\boldsymbol{k})A_{0}(\boldsymbol{p}-\boldsymbol{k}^{\prime})\phi(\boldsymbol{k})\phi(\boldsymbol{k}^{\prime})\cos[pb\hat{\boldsymbol{x}}\cdot(\hat{\boldsymbol{k}}-\hat{\boldsymbol{k}}^{\prime})]\\
 &  & \times[(E_{p}+m)^{2}+(E_{p}-m)^{2}\hat{\boldsymbol{k}}\cdot\hat{\boldsymbol{k}}^{\prime}-p^{2}\hat{\boldsymbol{p}}\cdot(\hat{\boldsymbol{k}}+\hat{\boldsymbol{k}}^{\prime})]
\end{eqnarray*}

For pseudoscalar potential, we have
\begin{eqnarray*}
\mathcal{P}_{1}^{PS}(b) & = & \frac{\alpha}{4(2\pi)^{6}}\int dpp^{6}\int d\Omega_{p}\int d\Omega d\Omega^{\prime}A_{0}(\boldsymbol{p}-\boldsymbol{k})A_{0}(\boldsymbol{p}-\boldsymbol{k}^{\prime})\phi(\boldsymbol{k})\phi(\boldsymbol{k}^{\prime})\cos[pb\hat{\boldsymbol{x}}\cdot(\hat{\boldsymbol{k}}-\hat{\boldsymbol{k}}^{\prime})]\\
 &  & \times\{\hat{\boldsymbol{y}}\cdot(\boldsymbol{k}\times\boldsymbol{k}^{\prime})+\hat{\boldsymbol{y}}\cdot[\hat{\boldsymbol{p}}\times(\hat{\boldsymbol{k}}-\hat{\boldsymbol{k}}^{\prime})]\}
\end{eqnarray*}
\begin{eqnarray*}
\mathcal{P}_{0}^{PS}(b) & = & \frac{\alpha}{4(2\pi)^{6}}\int dpp^{6}\int d\Omega_{p}\int d\Omega d\Omega^{\prime}A_{0}(\boldsymbol{p}-\boldsymbol{k})A_{0}(\boldsymbol{p}-\boldsymbol{k}^{\prime})\phi(\boldsymbol{k})\phi(\boldsymbol{k}^{\prime})\sin[pb\hat{\boldsymbol{x}}\cdot(\hat{\boldsymbol{k}}-\hat{\boldsymbol{k}}^{\prime})]\\
 &  & \times(1+\hat{\boldsymbol{k}}\cdot\hat{\boldsymbol{k}}^{\prime}-\hat{\boldsymbol{p}}\cdot\hat{\boldsymbol{k}}^{\prime}-\hat{\boldsymbol{k}}\cdot\hat{\boldsymbol{p}})
\end{eqnarray*}

\bibliographystyle{apsrev}
\addcontentsline{toc}{section}{\refname}\bibliography{ref-20200628}

\end{document}